\documentstyle[12pt,epsf]{article}

\begin{document}
\def\beq{\begin{equation}}
\def\eeq{\end{equation}}
\def\beqa{\begin{eqnarray}}
\def\eeqa{\end{eqnarray}}
\def\noin{\noindent}
\def\grad{{\bf \nabla}}
\def\bv{{\bf v}}
\def\bB{{\bf B}}
\def\bJ{{\bf J}}
\def\bE{{\bf E}}
\def\pa{\partial}
\def\eps{\epsilon_{\alpha\beta}}

\titlepage
\begin{flushright} QMW-PH-95-45
\end{flushright}
\vspace{4ex}

\begin{center} \bf

{\bf\Large Conformal Models of Magnetohydrodynamic Turbulence}\\

\rm

\vspace{10ex}

OMDUTH COCEAL\footnote{email: coceal@qmw.ac.uk} and
STEVEN THOMAS\footnote{email: s.thomas@qmw.ac.uk}\\
\vspace{10ex}

{\it Department of Physics\\
Queen Mary and Westfield College\\
Mile End Road\\
London E1\\
U.K.}\\

\vspace{14ex}
ABSTRACT

\end{center}
\noindent
Following the previous work of Ferretti and Yang on the role of magnetic fields
in the
theory of conformal turbulence, we show that non-unitary minimal model
solutions to
2-dimensional magnetohydrodynamics (MHD) obtained by dimensional
reduction from 3-dimensions
exist under different (and more restrictive) conditions. From a
 3-dimensional point of view, these conditions are equivalent to
perpendicular flow, in which the magnetic and velocity fields are
orthogonal.
 We also extend the analysis
to the finite conductivity case and present some approximate solutions, whose
connection to the exact ones of the infinite conductivity case is also
discussed.

\newpage
\section{Introduction}
In recent years it has become clear that the
proposal by Polyakov [1,2], that turbulent flow in two spatial
dimensions may be understood in terms of certain non-unitary conformal
field theories (CFT) (at least in the inviscid limit), has provided new
insights into
the problem of 2-d turbulence in general.
A number of authors extended this work, and found minimal model solutions
under a variety of different conditions [3-6].  In addition,  generalizations
to
flows in the presence of boundaries have been studied [7,8]. More recently,
there have been proposals concerning possible CFT solutions to turbulence
even when viscocity is present [9]. There are also results concerning
perturbations of the underlying CFT solutions
to 2-d  turbulence where the perturbations are either 2-dimensional in origin
[10]
or three-dimensional [11].

Despite the elegance of the CFT approach, one of  its least attractive aspects
is  the
appearance of an infinite number of possible solutions (the inviscid case).
The existence of an infinite number of solutions is an indication that not all
the
physically relevant constraints are being imposed. In an effort to understand
these issues
further, it is useful to study  a more complex turbulent system  than that of
an ordinary fluid
in 2-dimensions. To this end, Ferretti and Yang [13] studied CFT solutions to
an
effectively 2-dimensional  theory of
{\em ideal}  ( inviscid, infinite conductivity)
magnetohydrodynamical turbulence obtained by dimensional reduction from 3
dimensions.
 Here there is at least a possibility of  obtaining
more stringent conditions than those found in ordinary fluids.  The single
Navier-Stokes equation  of the simple fluid is replaced by 4 coupled
equations, so that the
resulting Hopf equations are rather more complicated. However, the authors of
[13]
considered a self-consistent truncation of these equations, by setting  $B \ =
V \ = 0$,
where $B$ and $V$ are scalars interpreted as third components of the magnetic
and
velocity fields of a  3-dimensional plasma.\footnote{ This limit defines pure
MHD in 2-dimensions.}
 In this limit, the CFT solutions obtained are
different from those of ordinary Polyakov turbulence, but as found in [14] it
is still apparent
that one can easily generate a large (possibly infinite) number of such
solutions via
non-unitary minimal models. So in this respect, we are no closer to finding a `
less dense  '
set of solutions.

In this paper amongst other things we shall also consider CFT solutions to
effective 2-d  ideal
magnetohydrodynamical turbulence, but with a different self-consistent
truncation than
the one considered in [13]. Our choice is physically equivalent to so-called `
perpendicular '
flow in 3-dimensions,  in which the magnetic and velocity fields are
constrained to be
at right angles to one another. We find that the corresponding constraints on
CFT solutions
imply that they should be a subset of those found in usual Polyakov turbulence,
and
furthermore  they
are very much more stringent than those discussed previously in the literature.
Indeed, in the first 80 or so non-unitary minimal model solutions
describing turbulence with either constant enstrophy or energy [3],  only 3 of
these correspond
to  solutions of  MHD turbulence in the limit mentioned.  As well as {\em
ideal} 2-d MHD , we also
consider the situation with finite conductivity $\sigma$.  Although this
necessarily introduces
dissipative effects,  we show that  it is possible to define a generalized
vorticity which
satisfies the steady Hopf equations, and upon which one can impose constant
flux. The conditions
in this case are so restrictive that thus far we have only found approximate
solutions  after
searching through all non-unitary minimal models with $q < 500 $.

The structure of the paper is as follows. In section 2, we discuss the
equations of 3-d MHD when dimensionally reduced to 2-d, and   briefly review
the results
of [13]. We then analyze CFT solutions under the perpendicular flow conditions
mentioned
above, when  a variety of different flux constraints are imposed, and exhibit a
number of
exact solutions. In section 3, the Hopf equations of 2-d MHD turbulence with
finite conductivity
are studied together with  flux constraints, and approximate solutions are
given. We conclude
in section 4 with some comments concerning the relation between the exact and
approximate
solutions found.
\newpage
\section{Minimal model solutions of 2-D MHD Turbulence}
It is well known (see for example [12])
that an approximate description of 3-dimensional plasmas
is given by Magnetohydrodynamic (MHD) equations,
which combine Maxwell's equations with those of hydrodynamics under the
simplifying assumptions of low frequency, low temperature and infinite
conductivity. Under the further assumption of incompressibility, the
{\em ideal} MHD equations take the form [12]

\beqa\label{eq:2.1}
\grad\cdot\bv & = & 0 \nonumber \\
\grad\cdot\bB & = & 0 \nonumber \\
\pa_{t}\bv + \bv\cdot\grad\bv & = & -\frac{1}{\rho_{m}}\grad P +
\frac{1}{c\rho_{m}}\bJ
\times\bB \\
\grad\times(\bv\times\bB) & = & \pa_{t}\bB \nonumber \\
\grad\times\bB & = & \frac{4\pi}{c} \bJ \nonumber
\eeqa

\noin
Here $\bv$ is the velocity field, $\bB$ is the magnetic field, $\bJ$ the
electric current, $P$ the pressure, and $\rho_{m}$ the mass density of
the plasma.
We can use the last equation to substitute for $\bJ$ in the third (the
Navier-Stokes equation). Then after taking the curl of the Navier-Stokes
equation to get rid of the pressure term, the MHD equations become:
\beqa\label{eq:2.2}
\grad\cdot\bv & = & 0 \nonumber \\
\grad\cdot\bB & = & 0 \nonumber \\
\grad\times(\pa_{t}\bv + \bv\cdot\grad\bv) & = &
\frac{1}{4\pi\rho_{m}}\grad\times
(\bB\cdot\grad\bB) \\
(\bB\cdot\grad)\bv - (\bv\cdot\grad)\bB & = & \pa_{t}\bB \nonumber
\eeqa
In this form the MHD equations display some evident symmetry between the $\bv$
and $\bB$ fields. It is therefore reasonable to expect that $\bB$ will play a
similar role to $\bv$ in some generalized MHD conformal turbulence scenario.
This observation was exploited by Ferretti and Yang [ 13]. First they reduced
the theory to an effectively two-dimensional one by requiring that all fields
be independent of the $z$-coordinate:
\beq\label{eq:2.3}
\pa_{3}\bv = 0,~~~\pa_{3}\bB = 0
\eeq
Equation (\ref{eq:2.3}) then implies that the first two components of $\bv$ and
$\bB$ can be written as
\beq\label{eq:2.4}
B_{\alpha} = \epsilon_{\alpha\beta}\pa_{\beta}A,~~~v_{\alpha} = \epsilon_
{\alpha\beta}\pa_{\beta}\psi,~~~~~ \alpha,\beta=1,2.
\eeq
The third components $B_{3}\equiv B$ and $v_{3}\equiv V$, together with the
stream function $\psi$ and magnetic potential $A$ are all two-dimensional
scalars.

\noin
If we now define, in analogy with vorticity $\omega$, the ``magnetic vorticity"
\beq\label{eq:2.5}
\Omega\equiv \epsilon_{\alpha\beta}\pa_{\alpha}B_{\beta}=-\pa_{\alpha}
\pa_{\alpha}A,
\eeq
and the two-dimensional operator
\beq\label{eq:2.6}
{\cal A}\equiv \eps\pa_{\beta}A\pa_{\alpha}
\eeq
the 2-D MHD equations take the form [13]:
\beqa\label{eq:2.7}
\dot{\omega} + \eps\pa_{\beta}\psi\pa_{\alpha}\omega & = &
\frac{1}{4\pi\rho_{m}}{\cal A}\Omega \nonumber \\
\dot{A} + \eps\pa_{\beta}\psi\pa_{\alpha}A & = & 0 \nonumber \\
\dot{V} + \eps\pa_{\beta}\psi\pa_{\alpha}V & = & \frac{1}{4\pi\rho_{m}}{\cal
A}B
\\
\dot{B} + \eps\pa_{\beta}\psi\pa_{\alpha}B & = & {\cal A}V
\nonumber
\eeqa
It is evident from eqn (\ref{eq:2.7}) that $V$ and $B$ are absent from the
equations for  $\dot{\omega}$ and $\dot{A}$. Hence, one can set $B$ and $V$ to
zero self-consistently, and study a simplified set of equations.

We therefore begin by examining the meaning and possible solution of the
first two equations in (\ref{eq:2.7}). Ferretti and Yang [13] pointed out that
the additional term
on the right hand side of the first equation destroys enstrophy conservation.
However, the second equation implies that the analogous quantity
\beq\label{eq:2.8}
G=\frac{1}{2}\int~A^2~d^2 x
\eeq
is now conserved. One can also check that total
energy
\beq\label{eq:2.9}
E=\frac{1}{2}\int~(v_{\alpha}v_{\alpha} + B_{\alpha}B_{\alpha})~d^2 x
\eeq
is a constant of the motion.
Proceeding in analogy with ordinary Polyakov (hydrodynamic) turbulence [1,2],
the authors of [13] interpreted $\psi$ and $A$ as primary fields in some
CFT and proceeded to derive constraints on such
 solutions by the requirements of steady solutions to the
inviscid Hopf equations, i.e.
\beq\label{eq:2.10}
\dot{A}=0
\eeq
and constancy of $A$-flux on scales $r \sim L $
\beq\label{eq:2.11}
<\dot{A}(r)A(0)>~\approx~r^0.
\eeq
$L$ being an infrared (IR) cutoff whose value is typically the size of the
largest
coherent motion in the system.
\noin
Beginning with the second equation in (\ref{eq:2.7})
\beq\label{eq:2.12}
\dot{A}=-\eps\pa_{\beta}\psi\pa_{\alpha}A
\eeq
point-split regularization gives:
\beq\label{eq:2.13}
\dot{A}~\sim~|a|^{2(\Delta_{\chi}-\Delta_{\psi}-\Delta_{A}+1)}[L_{-2}\bar{L}_
{-1}^{2}-\bar{L}_{-2}L_{-1}^{2}]~\chi,
\eeq
where $\chi$ is the minimal dimension field in the OPE of $\psi$ with $A$.
This fixes the dimension of $\dot{A}$ as $2+\Delta_{\chi}$.
Thus the Hopf equation for $A$ implies
\beq\label{eq:2.14}
\Delta_{\chi}~>~\Delta_{\psi}+\Delta_{A}-1
\eeq
while constancy of $A$-flux gives
\beq\label{eq:2.15}
\Delta_{\chi}+\Delta_{A}=-2
\eeq

\noin
Next we consider the relevance of the first equation in (\ref{eq:2.7}) in the
CFT
 context. The basic idea is that stationary turbulence implies
 vanishing of $\dot{\omega}$ in the inviscid
limit, with $\dot{\omega } $ defined through the Navier-stokes
equation for $\omega $, (\ref{eq:2.7}). In the point-splitting scheme, this is
implemented by the limit
$a\rightarrow 0$, where $a$ is the usual ultraviolet cutoff, proportional to
the
viscosity.
The first equation can be interpreted as a definition of $\dot{\omega}$:
\beq\label{eq:2.16}
\dot{\omega} = \eps\pa_{\beta}\psi\pa_{\alpha}\pa^{2}\psi - \frac{1}{4\pi
\rho_{m}}\eps\pa_{\beta}A\pa_{\alpha}\pa^{2}A
\eeq
After performing the regularization, operator product expansions (OPE's),
and differentiations, we obtain
\beq\label{eq:2.17}
\dot{\omega}=|a|^{2(\Delta_{\varphi}-2\Delta_{A})}{\cal L}\varphi - |a|^{2
(\Delta_{\phi}-2\Delta_{\psi})}{\cal L}\phi = 0,
\eeq
where in eqn (\ref{eq:2.17}), ${\cal
L}=[L_{-2}\bar{L}_{-1}^{2}-\bar{L}_{-2}L_{-1}^{2}]$
and $A \times A=[\varphi ]+\cdots,$$ $$\psi \times \psi=[\phi]+\cdots.$
where $\varphi $ and $\phi $ are minimal dimension fields.
The Hopf equation for $\omega$  thus
dictates two Hopf conditions for $\psi$ and $A$ respectively:
\beq\label{eq:2.18}
\Delta_{\varphi}>2\Delta_{A},~~~\Delta_{\phi}>2\Delta_{\psi}
\eeq
These are then the two additional constraints coming from the first equation in
(\ref{eq:2.7})
that any prospective CFT solution of 2-D MHD must satisfy. This brings the
total number of constraints to four (three Hopf conditions and one constant
flux constraint).
\noin
As far as the conformal dimension of $\dot{\omega}$ is concerned, it is
determined in the limit $a \rightarrow 0$ by whichever is the lower dimension
field, $\phi_{m}$, between the minimal dimension fields $\phi$ and $\varphi$ in
the OPE
$\psi\times \psi$ and $A\times A$ respectively:
\beq\label{eq:2.19}
\Delta_{\dot{\omega}}=2+\Delta_{\phi_{m}}
\eeq

The simplest solution found by Ferretti and Yang is the (2,13) non-unitary
minimal model,
which comprises six primary fields $\psi_{1,1}$ to $\psi_{1,6}$.
Interestingly, there is a one-to-one correlation between these and the six
fields $I$, $\psi$, $\phi$, $A$, $\chi$ and $\varphi$ respectively.
As an immediate physical consequence, one can compute the kinetic and magnetic
energy spectra. These are given by
\beq\label{eq:2.20}
E_{k} \sim k^{1+4\Delta_{\psi}},~~~E_{m} \sim k^{1+4\Delta_{A}}.
\eeq
For the (2,13) model, we have $\Delta_{\psi}=-5/13$ and $\Delta_{A}=-12/13$,
giving
\beq\label{eq:2.21}
E_{k} \sim k^{-7/13},~~~E_{m} \sim k^{-35/13}.
\eeq
It is not difficult to find more minimal models satisfying the above
constraints. Indeed, because the latter are not very stringent, one encounters
a  proliferation of solutions with a correspondingly wide range of
spectra (see Rahimi et al [14]). The situation here is somewhat worse than that
of
Polyakov turbulence; for example there are five models of the form $(2,2n+1)$:
$(2,13), (2,17), (2,19), (2,23)$ and $(2,27)$ with completely different spectra
predictions in each case. It seems evident, therefore, that some more
stringent constraint is to be sought that would limit the number of possible
solutions and make definite predictions regarding the spectra.

As a step towards placing further restrictions on solutions of
MHD turbulence we shall consider a
special limit of the MHD equations and derive additional constraints on
possible CFT solutions.
We set $A=V=0$ in the equations of motion (\ref{eq:2.7}). Note that this
implies that the
fields $\bv$ and $\bB$ are of the form $\bv=(v^1,v^2,0)$ and $\bB=(0,0,B)$,
i.e. the magnetic field is perpendicular to the plane of fluid flow.
These so called `perpendicular flow ' conditions are familiar in the study
of 3-dimensional MHD [15]. In the 2-dimensional case,
the resulting equations of motion then take the particularly simple and
symmetric form:
\beq\label{eq:2.22}
\pa_{t}\omega + \eps(\pa_{\beta}\psi)(\pa_{\alpha}\omega) = 0
\eeq
\beq\label{eq:2.23}
\pa_{t}B + \eps(\pa_{\beta}\psi)(\pa_{\alpha}B) = 0
\eeq
Eqn (\ref{eq:2.22}) is just the usual inviscid Navier-Stokes equation of
hydrodynamic turbulence whilst eqn (\ref{eq:2.23})  is its magnetic analogue.
Hence any solution of perpendicular flow MHD has to satisfy the usual Polyakov
constraints. We further note that eqn (\ref{eq:2.22}) is exactly of the same
form as the equation for $A$, with $A$ being replaced by $B$. Hence they
have the same solutions. We therefore conclude that, as would perhaps be
expected, our solutions will form a subset of both the hydrodynamic set of
solutions and magnetic ones mentioned above. This is a rather stringent
requirement and indeed we found that solutions are somewhat scarce. We can
categorize these in terms of different possible (and mutually exclusive)
constant flux constraints: constant enstrophy and constant energy,
(the latter allows for energy cascades from small to large scales as
envisaged by Kraichnan [16], Leith [17] and Batchelor [18]). This arises
from the observation that enstrophy and kinetic energy can now both be
conserved
(since $B_{\alpha}=0$). We also now have a third conserved quantity, the
integral of the square of the magnetic field $B$. This allows us to impose a
constant $B$-flux constraint as well, leading to the following possibilities:

\vskip0.2in
{\bf (a) Constant enstrophy flux + constant $B$-flux}
\beqa\label{eq:2.24}
\Delta_{\psi} + \Delta_{\phi} & = & -3 \nonumber \\
\Delta_{B} + \Delta_{\chi} & = & -2
\eeqa

\vskip0.2in
{\bf (b) Constant energy flux + constant $B$-flux}
\beqa\label{eq:2.25}
\Delta_{\psi} + \Delta_{\phi} & = & -2 \nonumber \\
\Delta_{B} + \Delta_{\chi} & = & -2
\eeqa
where $\psi\times\psi=[\phi]+\cdots, \quad B\times\psi=[\chi]+\cdots$
In either case the Hopf conditions~ $\dot{\omega}=0$ and $\dot{B}=0$~ further
require that,

\beqa\label{eq:2.26}
\Delta_{\phi} & > & 2\Delta_{\psi} \\
\Delta_{\chi} & > & \Delta_{B} + \Delta_{\psi} - 1
\eeqa
We note that the second inequality, eqn (\ref{eq:2.27}), implies together with
the $B$-flux
constraint eqn (\ref{eq:2.25}), that $\Delta_{\chi}>-3/2 + \Delta_{\psi}/2$.

\noin
These are not the only possibilities. Cateau et al [6] also discuss the
so-called Saffman (or discontinuity of vorticity) constraint, which has been
proposed as an alternative to the constant enstrophy or constant energy
conditions. In this case the conformal dimensions $\Delta_{\psi} $
and $\Delta_{\phi} $ satisfy
\beq\label{eq:2.28}
5\Delta_{\psi}+\Delta_{\phi}+9=0
\eeq
We surveyed Polyakov models in the range $q\leq500$ for constant enstrophy
(44 models) and for constant energy (41 models) as listed in the paper by
Lowe [3]. We also considered
the Saffman solutions given in Cateau et al [6] (9 models).
\vskip0.2in
\noindent
{\bf Constant enstrophy solutions}
\vskip0.2in
\noin
We only found one solution in this category in the range considered,
the $(21,166)$ model.
This is a large model with 1650 primary fields! The identifications are as
follows:
$ \psi\equiv\psi_{4,31}, \Delta_{\psi}\approx-1.496, ~
\phi\equiv\psi_{7,55}, \Delta_{\phi}\approx-1.504 $
and
$ B\equiv\psi_{7,61}, \Delta_{B}\approx-0.492, ~
\chi\equiv\psi_{10,79}, \Delta_{\chi}\approx-1.508.$
Here $\chi$ is actually the minimal dimension field in the model.
\vskip0.2in
\noin
{\bf Constant energy solutions}
\vskip0.2in
\noin
In this case we note an interesting possibility, namely, if we take a
solution of the constant energy constraint equation and identify $B$ with
$\phi$ and $\chi$ with $\psi$ then we obtain a simultaneous solution of the
second constraint equation. With this identification the Hopf condition
$\Delta_{\chi}>\Delta_{B}+\Delta_{\psi}-1$ implies that $\Delta_{\phi}<1$,
which is automatically satisfied. However we still have to fulfill the
requirement that $\chi$ (or $\psi$) is the minimal dimension field in the OPE
of $B$ with $\psi$ (ie $\phi$ with $\psi$). This is a non-trivial condition,
and it seriously limits the possible solutions. Amongst all of the  constant
energy minimal model solutions $ (p,q) $ with  $q < 500 $
 (see Lowe [3])  only $(10,59)$  and $(59,344) $ satisfy this. In the first
case we
have
$\psi~\equiv~\psi_{1,6},~\phi~\equiv~\psi_{1,5}$ and in the second,
$\psi~\equiv~\psi_{6,135},~\phi~\equiv~\psi_{5,29}$ . In both cases we have
$B~\equiv~\phi,~\chi~\equiv~\psi $,
where $\psi$ (or $\chi$) is the lowest dimension operator in the model.
We also note that in both models $\psi$ and $\phi$ have roughly the same
dimension, each
being very close to 1. Indeed the OPE constraints tend to indicate that
this is a generic feature of any similar solutions.
No other solutions were found in the range investigated.
\vskip0.2in
\noin
{\bf Saffman solutions}
\vskip0.2in
\noin
Here things are somewhat easier. If we consider CFT solutions
where the Saffman conditions (eqn. (27)) are imposed on the enstrophy,
the very first model $(2,17)$ provides a
solution to the MHD case. This is a small model with only 8 fields.
The field content and identifications are given in Table 1. Here $\varphi$ is
the minimal field in the OPE $B\times B$.
Another solution was the $(5,46)$ model, with
$\psi~\equiv~\psi_{1,13},~\phi~\equiv~\psi_{1,9}$
and
$B~\equiv~\psi_{1,17}, $ \hfill
$\chi~\equiv~\psi_{1,9}.$ Thus we have found that two of the 9 models listed in
[6]
are also solutions to the corresponding MHD case. Compared to the previous
cases of
constant energy $\ $ enstrophy (where we only found 3 models out of a list of
about 80), conformal solutions to MHD turbulence with Saffman conditions
imposed are easier to find.

\def\strut{\vrule height 2ex depth 2ex width 0pt}
\begin{table}
\small
\centering
\begin{tabular}{|c||c|c|c|c|c|c|c|c|}
\hline
\strut Field    & $\psi_{1,1}$  & $\psi_{1,2}$ & $\psi_{1,3}$ & $\psi_{1,4}$ &
$\psi_{1,5}$ & $\psi_{1,6}$ & $\psi_{1,7}$ & $\psi_{1,8}$ \\
\hline
\strut Interpretation &    $I$    &    $B$       & $\varphi $    &    ?
&             &  $\psi$      &   $\chi$     &    $\phi$    \\
\hline
\strut $\Delta\approx$ &    $0$    &  $-0.41$       &    $-0.76$     &
$-1.05$    &    $-1.29$     &    $-1.47$     &   $-1.59$      &    $-1.65$
\\
\hline
\end{tabular}
\vskip0.2in
\caption{Primary fields and their identifications in $(2,17)$.}
\end{table}
\vskip0.2in
\noin
{\bf Energy spectra}
\vskip0.2in
\noin
One can easily work out the immediate physical consequence of these solutions,
namely the energy spectra. Here kinetic energy $1/2\int~(v_{\alpha}v_{\alpha})
{}~d^{2}x$ and magnetic energy  $\qquad\qquad$
$1/2 \int~B^2~d^{2}x$ are each conserved
independently. We define their respective spectra $E_{{\rm kin.}}(k)$ and
$E_{{\rm mag.}}(k)$ by:
\beq\label{eq:2.29}
\int~dk~E_{{\rm kin.}}(k)~=~\frac{1}{2} <v_{\alpha}v_{\alpha}>
\eeq
\beq\label{eq:2.30}
\int~dk~E_{{\rm mag.}}(k)~=~\frac{1}{2} <B~B>
\eeq
where $k=\sqrt{k_{\alpha}k_{\alpha}}$. This gives
\beq\label{eq:3.31}
E_{{\rm kin.}}(k)~\sim~k^{1+4\Delta_{\psi}}, \qquad\qquad
E_{{\rm mag.}}(k)~\sim~k^{-1+4\Delta_{B}}
\eeq

\noin
We summarize our results in Table 2, which gives the dimensions of
all the relevant fields as well as the spectra predictions of each model.
We note in particular the approximate Kolmogorov-like kinetic energy spectrum
of the
constant energy models. This is due to the dimension of $\psi$ being close
to 1 in these solutions, as remarked earlier. It is also of interest that the
spectra of the other models are roughly consistent with each other.
\def\strut{\vrule height 2ex depth 2ex width 0pt}
\vskip0.2in
\begin{table}
\small
\centering
\begin{tabular}{|c||c|c|c|c|c|}
\hline
\strut  Model   & $(21,166)$ & $(10,59)$ & $(59,344)$ & $(2,17)$ & $(5,46)$ \\
\hline\hline
\strut $\psi$ & $\Delta_{4,31}\approx-1.496$ & $\Delta_{1,6}\approx-1.017$
&$\Delta_{6,35}\approx-1.0049$ & $\Delta_{1,6}\approx-1.471$ &
$\Delta_{1,13}\approx-1.435$ \\
\hline
\strut $\phi$ & $\Delta_{7,55}\approx-1.504$ & $\Delta_{1,5}\approx-0.983$
&$\Delta_{5,29}\approx-0.99951$ &$\Delta_{1,9}\approx-1.647$ &
$\Delta_{1,9}\approx-1.826$ \\
\hline
\strut $B$ &    $\Delta_{7,61}\approx-0.492$ & $\Delta_{1,5}\approx-0.983$
&$\Delta_{5,29}\approx-0.99951$&$\Delta_{1,2}\approx-0.412$ &
$\Delta_{1,17}\approx-0.174$ \\
\hline
\strut $\chi$ & $\Delta_{10,79}\approx-1.508$ & $\Delta_{1,6}\approx-1.017$ &
$\Delta_{6,35}\approx-1.0049$ &$\Delta_{1,7}\approx-1.588$ &
$\Delta_{1,9}\approx-1.826$ \\
\hline
\strut $\Delta_{0}\approx$ & $-1.508$ & $-1.017$ &$ -1.0049$ &$-1.647$ &
$-1.826$ \\
\hline
\strut $1+4\Delta_{\psi}$ & $-4.98$ & $-3.07$ & $-3.02$&$-4.88$ & $-4.74$ \\
\hline
\strut $-1+4\Delta_{B}$ & $-2.97$ & $-4.93$ & $-5.00$ &$-2.65$ & $-1.70$ \\
\hline
\end{tabular}
\vskip0.2in
\caption{All exact minimal model solutions of perpendicular flow turbulent MHD
and their spectra, with $q < 500 $}
\end{table}

l

{\bf Approximate solutions}
\vskip0.2in
\noin
As a matter of interest, let us mention some approximate solutions of
conformal turbulence. They might be relevant if one cannot easily find an exact
solution, as in the finite conductivity case to be discussed below.
Falkovich et al [4] listed a number of minimal models with
a pair of fields satisfying the dimensional constraints of constant enstrophy
but ${\it not} $ the minimal field requirement of the OPE. We consider the
example of
the $(13,107)$ model for the purpose of illustration.
In this model, $\psi\equiv\psi_{5,39}$ and $\phi\equiv\psi_{4,34}$ with
$\Delta_{\psi}+\Delta_{\phi}=-3$. However, $\phi$ is not quite the minimal
field in the OPE $\psi\times\psi$, but has dimension very close to it.
Let us denote the true minimal field by $\varphi$. Thus,
\beq\label{eq:2.32}
\psi\times\psi\sim[\varphi ]+\cdots,~~~~~\Delta_{\varphi}>2\Delta_{\psi}.
\eeq
In the $a\rightarrow0$ limit, $\dot{\omega}$ becomes a level 2 field in
$[\varphi ]$ and therefore acquires dimension $2+\Delta_{\varphi}$.
The enstrophy flux is then given by
\beq\label{eq:2.33}
<\dot{\omega}\omega>~\sim~L^{\Delta},
\eeq
where
\beqa\label{eq:2.34}
\Delta & = & \Delta_{\dot{\omega}} + \Delta_{\omega}    \nonumber \\
       & = & (2+\Delta_{\varphi}) + (1+\Delta_{\psi}) \nonumber \\
       & = & \Delta_{\varphi} - \Delta_{\phi},
\eeqa
using $\Delta_{\psi}+\Delta_{\phi}=-3$ in the last line of eqn (\ref{eq:2.34}).
Hence, if $\Delta$ is sufficiently small, we can obtain a very good
approximation to constant enstrophy flux.

\noin
For the  $(13,107)$ model, we have
\beqa\label{eq:2.35}
\psi\times\psi & = & \psi_{5,39}\times\psi_{5,39} \nonumber \\
               &   &  \nonumber \\
               & = & \sum_{i=1}^{9}\sum_{j=1}^{77}\psi_{i,j} \nonumber \\
               &   &   \nonumber \\
               & = & [\psi_{5,41}]+\cdots
\eeqa
(Here the summations take place in increments of 2).
Hence $\varphi =\psi_{5,41}$, with $\Delta_{\varphi}\approx-1.587$. Moreover
$\Delta_{\phi}=\Delta_{4,34}\approx-1.553$ so $\Delta\approx-.03$.

\smallskip
\noin
One could also scan such solutions for approximate solutions of the further
$B$-flux constraint
\beq\label{eq:2.36}
\Delta_{B}+\Delta_{\chi}=-2,~~~B\times\psi\sim[\chi]+\cdots
\eeq

\noin
The $(13,107)$ model again furnishes a good example, with
$B=\psi_{5,35},~\Delta_{B}\approx-1.562$ and
$\chi=\psi_{4,32},~\Delta_{\chi}\approx-0.438.$
Checking the OPE $B\times\psi$, we find
\beqa\label{eq:2.37}
B\times\psi & = & \psi_{5,35}\times\psi_{5,39} \nonumber \\
               & & \nonumber \\
               & = & \sum_{i=1}^{9}\sum_{j=1}^{73}\psi_{i,j} \nonumber \\
               & & \nonumber \\
               & = & [\psi_{5,41}]+\cdots
\eeqa
(Again the summations are in step of 2).
Thus again $B\times\psi~\sim~[\varphi ].$

\noin
Repeating the above argument, we obtain for the $B$-flux
\beq\label{eq:2.38}
<\dot{B}B>~\sim~L^{\Delta'}
\eeq
where
\beqa\label{eq:2.39}
\Delta' & = & \Delta_{\dot{B}}+\Delta_{B} \nonumber \\
        & = & (2+\Delta_{\varphi})+\Delta_{B}
\eeqa
Then, using $(2+\Delta_{B})=-\Delta_{\chi}$, we obtain
\beqa\label{eq:2.40}
\Delta' & = & \Delta_{\varphi}-\Delta_{\chi} \nonumber \\
        & \approx & -0.02
\eeqa
\noin
In a similar spirit we can look for approximate solutions of the $B$-flux
constraint among the exact solutions of Polyakov turbulence.
We found a few such close solutions within the range of models investigated.
In the constant enstrophy category the first is the model $(25,234)$ with
$\psi\equiv
\psi_{9,79},~~\phi\equiv\psi_{11,103},~~B\equiv\psi_{11,95},~~\chi\equiv
\psi_{9,83}$ and $\Delta'\approx-0.04$. There were more such approximate
solutions for constant energy, with the first example being the low-lying
model $(8,47)$ with  $\psi\equiv\psi_{3,17},~~\phi\equiv\psi_{3,18},
{}~~B\equiv\phi,~~\chi\equiv\psi$ and $\Delta'\approx-0.02$.
For the Saffman condition we have the
solution $(6,53)$ with
$\psi\equiv\psi_{1,12},~~\phi\equiv\psi_{1,9},~~B\equiv\psi_{1,16},~~\chi
\equiv \psi_{1,8}$ and $\Delta'\approx-0.02$.

\newpage
\section{MHD with finite conductivity}


In this case the full set of the modified MHD equations is as follows [12]:
\beqa\label{eq:3.1}
\grad\cdot\bv & = & 0 \nonumber \\
\grad\cdot\bB & = & 0 \nonumber \\
\grad\cdot\bJ & = & 0 \nonumber \\
\grad\cdot\bE & = & \rho_{c} \nonumber \\
\pa_{t}\bv + \bv\cdot\grad\bv & = & -\frac{1}{\rho_{m}}\grad P +
       \frac{\rho_{c}}{\rho_{m}}\bE + \frac{1}{c\rho_{m}}\bJ\times\bB \\
\bJ & = & \sigma (\bE + \frac{1}{c}\bv\times\bB) \nonumber \\
\grad\times\bE & = & -\frac{1}{c}\pa_{t}\bB \nonumber \\
\grad\times\bB & = & \frac{4\pi}{c} \bJ \nonumber
\eeqa

\noin
Now we have an additional $\bE$ field, and new parameters $\sigma$
(conductivity) and $\rho_{c}$ (electric charge density).
After substituting for $\bJ$ as before and again taking the curl of the
Navier-Stokes equation to eliminate the pressure term, we obtain

\beqa\label{eq:3.2}
\grad\cdot\bv & = & 0 \nonumber \\
\grad\cdot\bB & = & 0 \nonumber \\
\grad\cdot\bE & = & \rho_{c} \nonumber \\
\grad\times (\pa_{t}\bv + \bv\cdot\grad\bv) & = & \frac{\rho_{c}}{\rho_{m}}
   \grad\times\bE + \frac{1}{c\rho_{m}}\grad\times(\bB\cdot\grad\bB) \\
\grad\times\bE & = & -\frac{1}{c}\pa_{t}\bB \nonumber \\
\grad\times\bB & = & \frac{4\pi\sigma}{c} (\bE + \frac{1}{c}\bv\times\bB)
\nonumber
\eeqa

\noin
We note that these equations correctly reduce to the set of eqns (\ref{eq:2.1})
and (\ref{eq:2.2}) in the infinite conductivity limit (where
$\sigma\rightarrow\infty$
and $\rho_{c}\rightarrow 0$).

\noin
Next, we again perform dimensional reduction, additionally constraining the
$\bE$ field such that
\beq\label{eq:3.3}
\pa_{3}\bE=0
\eeq
and denoting the two-dimensional scalar field $E_{3}$ by $E$. We further note
that even though the  two-dimensional quantity $\pa_{\alpha}E_{\alpha}$
 is non-vanishing, any vector in 2-dimensions may be decomposed into
curl-free and divergence-free parts
\beq\label{eq:3.4}
E_{\alpha}=\eps\pa_{\beta}C+\pa_{\alpha}D.
\eeq
After some algebraic manipulation one finally arrives at the following set of
equations for the finite-conductivity MHD equations in 2-D:

\beqa\label{eq:3.5}
\dot{\omega} + \eps\pa_{\beta}\psi\pa_{\alpha}\omega & = & \frac{1}{4\pi
  \rho_{m}}{\cal A}\Omega - \frac{\rho_{c}}{\rho_{m}}
 (\frac{c}{4\pi\sigma}\pa_{\alpha}\pa_{\alpha}B-\frac{1}{c}[\eps\pa_{\beta}\psi
 \pa_{\alpha}B-{\cal A}V])\nonumber \\
\dot{A} + \eps\pa_{\beta}\psi\pa_{\alpha}A & = & -\frac{c^2}{4\pi\sigma}\Omega
 + {\rm constant.} \nonumber \\
\dot{V} + \eps\pa_{\beta}\psi\pa_{\alpha}V & = & \frac{1}{4\pi\rho_{m}}
  {\cal A}B + \frac{\rho_{c}}{\rho_{m}}(\frac{c}{4\pi
  \sigma}\Omega-\frac{1}{c}{\cal A}\psi) \\
\dot{B} + \eps\pa_{\beta}\psi\pa_{\alpha}B & = & {\cal A}V
+\frac{c^2}{4\pi\sigma}\pa_{\alpha}\pa_{\alpha}B  \nonumber \\
\pa_{\alpha}\pa_{\alpha}D & = & \rho_{c} \nonumber
\eeqa
Again, we find that the infinite conductivity limit is consistent with the
previous ideal MHD equations.
One should note that, interestingly, the field $C$ drops out completely while
$D$ decouples from the dynamics. Thus, we still only have four independent
dynamical variables $\omega$, $A$, $B$ and $V$. However, the presence of
$\sigma$ has a
non-trivial effect on the equations. In particular, it is no longer obviously
consistent to set $B = V = 0 $, as we shall now discuss.
\vskip0.2in
\noin
{\bf Case (I):} $B=V=0$
\vskip0.2in
\noin
In this case the above equations reduce to
\beqa\label{eq:3.6}
\dot{\omega} + \eps\pa_{\beta}\psi\pa_{\alpha}\omega & = & \frac{1}{4\pi
  \rho_{m}}{\cal A}\Omega \nonumber \\
\dot{A} & = & [\frac{c^2}{4\pi\sigma}\pa_{\alpha}\pa_{\alpha}A+{\cal
A}\psi]+{\rm
  constant.} \\
\frac{c^2}{4\pi\sigma}\pa_{\alpha}\pa_{\alpha}A+{\cal A}\psi & = &0 \nonumber
\eeqa
The last equation in (\ref{eq:3.6}) is a constraint which
forces $A$ to have a trivial time-dependence
($\dot{A}=\lambda$, where $\lambda$ is a constant). Hence, defining
$A'=A-\lambda t$, we have
\beq\label{eq:3.7}
\dot{A}'=0
\eeq
In other words the magnetic field is static. Clearly quantities of the form
$\int d^{2}x A^{p}$ and  $\int d^{2}x (B_{\alpha}B_{\alpha})^{p}$, for
any positive integer $p$, will all be conserved.
The existence of  a  constraint on $A$ ( eqn (\ref{eq:3.6}))
 however, proves fatal to the conformal approach.
In field theory, point-splitting tells us that ${\cal A}\psi$ is a level 2
field in the conformal family $[\chi]$ of some minimal dimension field $\chi$.
On the other hand, the other term $\pa_{\alpha}\pa_{\alpha}A$ is in $[A]$ at
level 1. Also we note that while ${\cal A}$ is an antisymmetric operator,
$\pa_{\alpha}\pa_{\alpha}$ is symmetric. Thus parity considerations also rule
out a CFT solution to this problem. We interpret this as a limitation of the
conformal approach to turbulence rather than as a fact of underlying physical
significance.
\vskip0.2in
\noin
{\bf Case (II):} $A = V = 0$
\vskip0.2in
\noin
Next we consider the limit $A = V = 0$, as in the infinite conductivity case.
The $V$ and $A$ equations are then automatically satisfied if we
choose the constant to be zero in eqns. (\ref{eq:3.6}). Thus unlike case (I)
 we do not obtain constraint equations, which as we have seen are
 obstacles to CFT solutions. The remaining
equations of motion become
\beq\label{eq:3.8}
\dot{\omega} + \eps\pa_{\beta}\psi\pa_{\alpha}\omega = - \frac{\rho_{c}}{
\rho_{m}}(\frac{c}{4\pi\sigma}\pa_{\alpha}\pa_{\alpha}B-\frac{1}{c}\eps\pa_
{\beta}\psi\pa_{\alpha}B)
\eeq
\beq\label{eq:3.9}
\dot{B} + \eps\pa_{\beta}\psi\pa_{\alpha}B = \frac{c^2}{4\pi\sigma}\pa_
{\alpha}\pa_{\alpha}B
\eeq
The immediate observation here is that we now have non-trivial finite effects
due to the $\pa_{\alpha}\pa_{\alpha}B$ terms in both equations. Therefore
$\dot{\omega}$ and $\dot{B}$ do not vanish in the inviscid limit $a\rightarrow
0$. One can also check that in this case there are no `classical' quadratic
conserved quantities as in the infinite $\sigma$ case.
Indeed, generically the presence of finite $\sigma $ induces
decay in $B$ and $\omega $ as discussed earlier.
 At first sight this might seem fatal for conformal turbulence.
However, further thought reveals an interesting feature which we now describe.

Let us assume that we have some CFT in which we identify $\psi$ and $B$ with
suitable primary fields such that the OPE's $\psi\times\psi$ and $\psi\times
B$ each have a positive defect of dimensions as before. However, in this case
we do not insist on any other condition at this stage. Of course any previous
solution of the additional constraints of the infinite conductivity case would
automatically be a
suitable candidate. Having fixed $\psi$ and $B$ one can then use
eqns (\ref{eq:3.8}) and (\ref{eq:3.9})
 as a definition of $\dot{\omega}$ and $\dot{B}$ respectively, ie
\beq\label{eq:3.10}
\dot{\omega}\sim [\phi]_{2}+[\chi]_{2}+\frac{1}{\sigma}[B]_{1}
\eeq
\beq\label{eq:3.11}
\dot{B}\sim [\chi]_{2}+\frac{1}{\sigma}[B]_{1}
\eeq
where the subscripts on the r.h.s.  of eqns. (\ref{eq:3.10}) and
(\ref{eq:3.11})
 indicate the level of the
corresponding secondary conformal fields.
Now by the assumption of positive defect of dimensions, (eqns (\ref{eq:2.26})
)),
$[\phi]_{2}$ and $[\chi]_{2}$ both vanish by
construction. Thus we obtain
\beq\label{eq:3.12}
\dot{\omega}=\frac{c\rho_{c}}{4\pi\sigma\rho_{m}}\pa_{\alpha}\pa_{\alpha}B
\eeq
\beq\label{eq:3.13}
\dot{B}=-\frac{c^{2}}{4\pi\sigma}\pa_{\alpha}\pa_{\alpha}B
\eeq
Now we notice that if we define
$W\equiv\omega+\frac{\rho_{c}}{c\rho_{m}}B$
then
\beq\label{eq:3.14}
\dot{W}=0
\eeq
So we discover that although correlation functions of $\omega$ or $B$ cannot
simultaneously satisfy Hopf equations describing steady flow,
correlations involving $W$ only do satisfy
Hopf-like equations at the u.v. (short distance) level that are steady:
\beq\label{eq:3.15}
<\dot{W}(x_{1})W(x_{2})\cdots>+<W(x_{1})\dot{W}(x_{2})\cdots>+\cdots=0.
\eeq
In other words, there is still a remnant of `steady ' turbulence present, even
though
the effect of finite conductivity on the N-S equation is similar to that of
finite viscosity in that it destroys the turbulence. Here, because we
have a second equation for $B$ which has a similar finite conductivity
dependence, we can still salvage steady turbulence
in the variable $W$. We mention at this point that as well as  the $\sigma $
dependent terms, $[\chi ]_2 $ also drops out of the equation of
motion of  $W$. Thus in principle it is only necessary to require
$[\phi]_2 $ vanishes to obtain steady turbulence in $W$. However, solutions
under these conditions can never make sense in the
 $\sigma \rightarrow \infty  $ limit. Furthermore, even though we cannot
expect a CFT solution to the steady Hopf equation for $B$, we still have
to make sense of eqn (\ref{eq:3.14}), which means requiring
$\Delta_{\chi} > \Delta_{\psi} + \Delta_{B} -1$ (we do not consider
here the situation where $[ \chi ]_2 $ is a null field.) Hence in what follows
we will impose this condition.
\vskip0.2in
\noin
{\bf Flux constraints}
\vskip0.2in
\noin
Since $\dot{B}$ is non-zero, there are short-distance violations of constant
$B$-flux. Using
\beqa\label{eq:3.16}
\dot{B} & \sim & [\chi]_{2}+\frac{1}{\sigma}[B]_{1} \nonumber \\
        & = & \overline{\lim_{a\rightarrow 0}} \  \ {\rm const.}\
(a\bar{a})^{\Delta_{\chi}-\Delta_{\psi}-\Delta_{B}+1}
        {\cal L}\chi - \frac{c^{2}}{4\pi\sigma}\pa^{2}B
\eeqa
we can evaluate the correlator $< \dot{B} (r) B(0) > $. Taking
$\Delta_{\chi} > \Delta_{\psi} + \Delta_{B} -1 $ we find for scales
$r << L $
\beqa\label{eq:3.17}
<\dot{B}(r)B(0)> & \sim &  - \frac{c^{2}}{4\pi\sigma}<\pa^{2}B(r)
   B(0)> \nonumber \\
   & \sim &  - \frac{c^{2}}{4\pi\sigma}|r|^{-2(1+2
   \Delta_{B})}
\eeqa
Clearly, finite $\sigma$ violates constant $B$-flux (as expected),
even on scales $r << L $. We note
that, depending on whether $\Delta_{B}<-1/2$ or $>-1/2$, this decay of $B$-flux
will occur quicker at longer or shorter lengthscales. Similar arguments
apply to enstrophy.

We next consider $W$-flux. The relevant correlator is $<W(r)\dot{W}(0)>$. From
the definition of $W$ we can evaluate this correlator as
\beqa\label{eq:3.18}
<W(r)\dot{W}(0)> = <\omega(r)\dot{\omega}(0)> + \frac{\rho_{c}}{c\rho_{m}}
 <\omega(r)\dot{B}(0)> + \frac{\rho_{c}}{c\rho_{m}}<B(r)\dot{\omega}(0)>
\nonumber \\
 + (\frac{\rho_{c}}{c\rho_{m}})^{2}<B(r)\dot{B}(0)>
\eeqa
There are no u.v. ($r << L $ ) contributions to the r.h.s. of
eqn (\ref{eq:3.18}) since $\dot{W}=0$ on these
scales. However as usual there could be infrared contributions on the
lengthscales $r \sim L$. Evaluating the various terms we find  a number of
cancellations occur leading to
\beq\label{eq:3.19}
<W(r)\dot{W}(0)>~ \sim~ -L^{-(2+\Delta_{\phi}+1+\Delta_{\psi})} -
\frac{\rho_{c}
 }{c\rho_{m}}L^{-(\Delta_{B}+\Delta_{\phi}+2)}
\eeq
At this stage we allow ourselves to speculate on the possibility of imposing
constant $W$-flux. Such a condition would require that
\beqa\label{eq:3.20}
\Delta_{\phi}+\Delta_{\psi}+3 & = & 0 \nonumber \\
\Delta_{\phi}+\Delta_{B}+2 & = & 0
\eeqa
The first condition in eqn (\ref{eq:3.20}) is
 the familiar constant enstrophy constraint. It follows naturally
by demanding constancy of $W$-flux as $\int W^{2} d^{2}x$ is some generalized
enstrophy ($W$ is the generalized vorticity). The second is similar to the
constant $B$-flux condition of the $\sigma$-infinite case, with $\chi$ being
the minimal field $\phi$ of $\psi\times\psi$, although
we should emphasize that here $B$ is not necessarily  constrained
by the additional requirement $B\times\psi\sim [\chi]$. Of course we also need
to satisfy the two Hopf conditions corresponding to the requirement that
$[\chi]_{2}$ and $[\phi]_{2}$ vanish in the inviscid limit.

Again these turn out to be  very stringent requirements for any prospective
solution
of finite $\sigma$ MHD.
The first constraint tells us that we are looking for a subset of the
constant enstrophy solutions of Polyakov turbulence,
which also contains  any field $B$ with dimension given by
\beq\label{eq:3.21}
\Delta_{B}=-2-\Delta_{\phi}
\eeq
We have checked all the constant enstrophy minimal models given in Lowe [3]
without any success.
One can however  find  approximate solutions. We list the first six of
them in Table 3, although we have found others
in our search through all minimal model solutions of turbulence with
$q < 500 $ which we have not listed. Indeed, it turns out to be easier to find
approximate
solutions in the case of finite $\sigma $ than in the infinite one,
 because as mentioned above
$B$ is not necessarily restricted to satisfy  the OPE $ B \times \psi \sim \chi
$.
  In table 3, $\Delta\equiv\Delta_{\phi}+\Delta_{B}+2$ is a measure
of the deviation from exact enstrophy flux constancy.

\def\strut{\vrule height 1ex depth 2ex width 0pt}

\begin{table}
\small
\centering
\begin{tabular}{|c||c|c|c|c|c|}
\hline
\strut  Model   & $\chi $ & $\phi$ & $\psi $ & $B$ & $\Delta$  \\
\hline\hline
\strut $(11,87)$& $ \Delta_{5,39} \approx-1.499 $& $\Delta_{3,23}\approx-1.492$
&$\Delta_{2,16} \approx-1.508 $ & $\Delta_{4,26}\approx-0.505$ & $0.003$ \\
\hline
\strut $(11,91)$ & $\Delta_{3,25}\approx-1.597$ & $\Delta_{3,25}\approx-1.597$
&$\Delta_{2,14}\approx-1.403$  &$\Delta_{4,26}\approx-0.409$ & $-0.007$ \\
\hline
\strut $(14,111)$ &$\Delta_{5,39}\approx-1.501$ & $\Delta_{1,7}\approx-1.486$
&$\Delta_{1,8}\approx-1.514$  &$\Delta_{5,34}\approx-0.510$ & $0.004$ \\
\hline
\strut $(14,115)$ &$\Delta_{4,34}\approx-1.544$  &$\Delta_{1,9}\approx-1.565$
&$\Delta_{1,6}\approx-1.435$ & $\Delta_{4,39}\approx-0.436$ & $-0.0008$ \\
\hline
\strut $(16,135)$ &$\Delta_{7,60}\approx-1.613$ & $\Delta_{7,59}\approx-1.639$
&$\Delta_{7,56}\approx-1.361$ & $\Delta_{1,15}\approx-0.363$ & $-0.002$ \\
\hline
\strut $(21,166)$ &$\Delta_{10,79}\approx-1.508$ &$\Delta_{7,55}\approx-1.504$
&$\Delta_{4,31}\approx-1.496$ &$\Delta_{7,61}\approx-0.492$ & $0.003$ \\
\hline
\end{tabular}
\vskip0.2in
\caption{Some approximate solutions of finite conductivity MHD.}
\end{table}
\vskip0.25in
Of particular noteworthiness is
the last model (21,166). This is also an exact solution of
the infinite conductivity case as we found earlier. Moreover, surprisingly,
the $B$-field here is precisely the same as before ($\psi_{7,61}$),
which is a consequence of $\Delta_{\chi}$ and
$\Delta_{\phi}$ being very close together. In this
case the deviation $\Delta$ is given by
\beqa\label{eq:3.22}
\Delta & = & \Delta_{B}+\Delta_{\phi}+2 \nonumber \\
       & = & (\Delta_{B}+\Delta_{\chi}+2)+(\Delta_{\phi}-\Delta_{\chi})
\nonumber \\
       & = & \Delta_{\phi}-\Delta_{\chi}
\eeqa
\newpage
\section{Conclusion}
First let us  make a few comments concerning the connection
between the approximate
 solutions discussed in the previous section, and the exact solutions of the
infinite $\sigma $
MHD turbulence with constant enstrophy obtained in section 2.
Clearly the limit $\sigma \rightarrow \infty $ is non-trivial in that the
second
 condition
in eqn (\ref{eq:3.20}) is replaced by the $B$-flux condition of eqn
(\ref{eq:2.24}),
and in addition,  $B$ is now restricted by the OPE $B \times \psi \sim \chi $.
All the approximate solutions in Table 3 except (21,166) fail to satisfy either
or both of
these conditions, so they remain only approximate solutions when $\sigma $ is
finite.
Remarkably, as we have already mentioned, (21,166) goes over to an exact
solution in the limit $\sigma \rightarrow \infty $. It is unfortunate that we
can
only verify this interesting property for the single exact solution found in
the range of
 models considered. Certainly such a property provides further motivation for
finding further
exact solutions to the constant enstrophy, $\sigma $ infinite  model of
magnetohydrodynamical turbulence considered in this paper.
Since it would seem very unlikely that the same minimal
model could describe these two limits in such a way by pure coincidence, we
are left with the open question of whether this might relate to some generic
feature of physical interest.

 As regards exact
solutions for $\sigma $ finite, we can only speculate at this point.
At first sight, it may  appear that there are  two possible outcomes as
 $\sigma \rightarrow \infty $,
either such solutions remain exact solutions or they do not. If they
remain exact, the simultaneous solutions of eqns (\ref{eq:2.24}) and
(\ref{eq:3.20} ) implies that $ \Delta_ {\phi}  \ = \ \Delta_{\chi} $ i.e.
$\phi $ is
identified with $\chi $.   Hence  $\chi $ (or $\phi $) would have to be the
minimal dimension
field in both the OPE $B \times \psi $ and $\psi \times \psi $, which is a very
restrictve
condition. If such solutions  do not remain exact, it would be interesting to
see
if they are at least approximate, because if this is the case then we have the
picture that
exact solutions with constant enstrophy and $\sigma $ infinite are
 approximate solutions when $\sigma $ is finite and vice versa.

In conclusion, we have shown that by requiring that CFT describe the special
case of perpendicular flow 3-d MHD
dimensionally reduced to 2-d, we have been led to impose an extra constraint
and thus drastically
cut down on the number of possible minimal models that can describe
this theory.
We showed that the perpendicular $B$-field produced a ``hydromagnetic
turbulence" effect superimposed upon the usual hydrodynamic turbulence in
some minimally interacting manner. Extending the analysis to the finite
conductivity case we discovered that the Hopf equation was still satisfied by
some generalized vorticity $W$, producing some mixed $\omega$ and $B$
turbulence
phenomenon. There are no analogous conserved quantities but the
prospect of $W$-flux constancy is not thereby ruled out. Some interesting
instances of minimal models offering approximate realizations of the latter
were identified and the issue of the relationship with the infinite
conductivity limit was raised.
\vskip1in
\begin{center}
{\bf\Large Acknowledgements}
\vskip0.2in
S.T would like to thank the Royal Society of Great Britain for financial
support.
\end{center}
\newpage
\vspace{2ex}
\begin{center}
\noin{\bf\Large References}
\end{center}
\vspace{2ex}
\begin{description}
\item{[1]} A .M. Polyakov, Princeton Univ. Preprint PUPT-1341, hep-th/9209046.
\item{[2]} A .M. Polyakov {\it Nucl. Phys.} {\bf B396} (1993) 367.
\item{[3]} D .A. Lowe {\it Mod. Phys. Lett.} {\bf A8} (1993) 923.
\item{[4]} G. Falkovich and A. Hanany, `Spectra of Conformal Turbulence ',
Preprint WIS-92-88-PH, hep-th/9212015 (1992).
\item{[5]} Y. Matsuo {\it Mod. Phys. Lett.} {\bf A8} (1993) 619.
\item{[6]} H. Cateau, Y. Matsuo and M. Umeski,  `Predictions on two-dimensional
turbulence by conformal field theory ', Preprint UT-652, hep-th/9310056 (1993).
\item{[7]} B.K. Chung, S. Nam, Q-H. Park and H.J. Shin, {\it Phys. Lett.} {\bf
B309}
(1993) 58.
\item{[8} B.K. Chung, S. Nam, Q-H. Park and H.J. Shin, {\it Phys. Lett.} {\bf
B317} (1993) 92.
\item{[9]} Ph. Brax, `The Coulomb Gas Behaviour of Two Dimensional Turbulence',
 Preprint SPhT 95/58, DAMTP 95/25 (1995).
\item{[10]} O. Coceal and S.Thomas, {\it Mod. Phys. Lett.}{\bf  A10} (1995)
2427.
\item{[11]} L. Moriconi,  `3-D Perturbations in Conformal Turbulence ' Preprint
PUPT-95- hep-th/9508040 .
\item{[12]} L. C. Woods  `Principles of Magnetoplasma Dynamics ',
Oxford Science publs. (1987).
\item{[13]} G. Ferretti and Z. Yang, {\it Europhys. Lett.} {\bf 22}  (1993)
639.
\item{[14]} M. R. Rahimi Tabar and S. Rouhani, `Turbulent Two Dimensional
Magnetohydrodynamics and Conformal Field Theory', Preprint IPM-94-67-REV,
hep-th/9503005.
\item{[15]} T. J. M. Boyd and J. J. Sanderson, ` Plasma Dynamics ', Nelson
(1969).
\item{[16]} R.H. Kraichnan, `Inertial ranges in two-dimensional turbulence',
{\it Phys. Fluids } {\bf 10} (1967) 1417.
\item{[17]} C.E. Leith,`Diffusion approximation in two dimensional turbulence',
{\it Phys. Fluids} {\bf 11} (1968) 671.
\item{[18]} G.K. Batchelor, `Computation of the energy spectrum in homogenuos
two-
dimensional turbulence', {\it Phys. Fluids Suppl. II} {\bf 12} (1969) 233.
\end{description}



\end{document}